\documentclass[aps,preprint,onecolumn,showpacs,amsmath,amssymb]{revtex4}
\usepackage{graphics}
\usepackage{multirow}

\newcommand{\element}[3]{\langle #1|#2|#3\rangle}

\begin{document}

\title{Decoherence due to an excited state quantum phase transition in
a two-level boson model}

\author{P. P\'erez-Fern\'andez$^{1}$, A. Rela\~{n}o$^{2}$,
  J. M. Arias$^{1}$, J. Dukelsky$^{2}$, and
  J. E. Garc\'{\i}a-Ramos$^{3}$}

 \affiliation{$^{1}$ Departamento de F\'{\i}sica At\'omica, Molecular y
  Nuclear, Facultad de F\'{\i}sica, Universidad de Sevilla,
  Apartado~1065, 41080 Sevilla, Spain \\
$^{2}$ Instituto de Estructura de la Materia, CSIC, Serrano 123,
E-28006 Madrid, Spain \\
$^{3}$ Departamento de F\'{\i}sica Aplicada, Universidad de Huelva, 21071 Huelva, Spain}

\email{armando@iem.cfmac.csic.es}
\date{\today}
\begin{abstract}
  The decoherence induced on a single qubit by its interaction with
  the environment is studied. The environment is modelled as a scalar
  two-level boson system that can go through either first order or
  continuous excited state quantum phase transitions, depending on the
  values of the control parameters. A mean field method based on the
  Tamm-Damkoff approximation is worked out in order to understand the
  observed behaviour of the decoherence. Only the continuous excited
  state phase transition produces a noticeable effect in the
  decoherence of the qubit. This is maximal when the
  system-environment coupling brings the environment to the critical
  point for the continuous phase transition. In this situation, the
  decoherence factor (or the fidelity) goes to zero with a finite size
  scaling power law.
\end{abstract}

\maketitle

\section{Introduction}

Decoherence is the quantum phenomenon by which the coherence of a
quantum system can be destroyed when it is put in contact with a large
environment \cite{Zurek:03,Sch:04}. The Schr\"oedinger equation is a
linear differential equation, consequently any linear combination of
solutions is also a solution of the problem. Thus, a general possible
quantum state is a superposition of quantum states. Nevertheless, such
a state does not appear in the classical macroscopic world. The
decoherence interpretation of quantum mechanics \cite{Zurek:03} claims
that this is due to the interaction with the environment, which
destroys the quantum correlations between the states of the system,
making it to transite from a quantum superposition state to a
classical-like mixture of states. Moreover, only a small set of states
take part of the classical-like mixture; they are called {\it pointer
  states} \cite{Zurek:03}.

The study of decoherence is important for several reasons: i) it might
be responsible for the emergence of classical properties out of the
underlying quantum nature of the physical systems, ii) it is a major
problem for the construction of a quantum computer since it will
produce the loss of the necessary quantum entanglement. Thus, both for
fundamental reasons (i) and for practical purposes (ii) it is
important to characterize the decoherence process and its effects on
the physical properties of a quantum system.

Along this line of study, it is important to address the issue of the
effect produced in the coherence of a quantum state when the
environment evolves between different quantum phases. There have been
several works on the relation between decoherence and an environmental
quantum phase transition
\cite{Paz:07,Paz:08,Rossini:07,Wang:08,Camalet:07,Yuan:07}. Recently,
we have presented a novel phenomenon in which the decoherence of the
system suffers drammatic changes when the environment crosses an
excited state quantum phase transition (ESQPT)\cite{Relano:08}. An
ESQPT is a nonanalytic evolution of the system as the control
parameters in the Hamiltonian vary. It is similar to a ground state
quantum phase transition but affecting to excited
states. Correspondingly, an ESQPT can be classified in the
thermodynamic limit as first order, when a crossing between two
excited levels is present, or continuous, when the number of
interacting levels is locally very large at an excited energy but
without crossings. 

In Ref. \cite{Relano:08} we presented briefly the case of a qubit in
interaction with an environment modelled as a two-level boson system
undergoing a continuous ESQPT. We used a particular simple Hamiltonian
in terms of single control parameter to model the environment in order
to show the main effect. Here we present a more extensive study of a
similar system including both first and second order ESQPT, and more
general sets of parameters. Together with the exact evolution of the
system, we present a simple mean field treatment. We show that the
decoherence is maximal when the interaction of the system with the
environment produces second order ESQPT, while no noticeable effects
are observed in the case of a first order ESQPT. For the former case,
a finite-size scaling analysis allows us to postulate that the
fidelity goes to zero as soon as the interaction between system and
environment is switched on. We also show that mean field treatment
provides a good description for the decoherence of the small system,
except around the critical points.

The paper is structured as follows. In Sect. II, we present the model
for the environment and study the phase diagram and its relation with
the density of energy levels. We then discuss the interaction of the
environment with the system. In Sect. III we show results for the
decoherence factor. Both exact numerical results for large boson
number, and an analytic mean field method with simple extensions of
the Tamm-Dankoff approximation are presented. In Sect. IV,
results for the decoherence factor in the case of continuous and first
order ESQPT, including a finite size scaling study for the decoherence
factor (or fidelity), are discussed. Finally, in Sect. V we summarize
giving the main conclusions of this work.

\section{The Model}

Following \cite{Paz:07}, we will consider our system composed by a
spin $1/2$ particle coupled to a spin environment by the Hamiltonian
$H_{\mathcal{SE}}$,
\begin{equation}
\label{HSE}
H_{\mathcal{SE}} = I_{\mathcal{S}} \otimes H_{\mathcal{E}} +
\left|0\right> \left<0\right| \otimes H_{\lambda_0} + \left|1\right>
\left<1\right| \otimes
H_{\lambda_1},
\end{equation}
\noindent where $\left|0\right>$ and $\left|1\right>$ are the two
components of the spin $1/2$ system, and $\lambda_{0}$, $\lambda _1$
the couplings of each component to the environment. The three terms
$H_{\mathcal{E}}$, $H_{\lambda_0}$ and $H_{\lambda_1}$ act on the
Hilbert space of the environment; therefore, it evolves with an
effective Hamiltonian depending on the state of the central spin
$H_{i}=H_{\mathcal{E}}+H_{\lambda _i}$, $i=0, 1$. The term
$H_{\lambda_i}$ makes it possible that the environment crosses a
critical point as a consequence of the interaction with the central
spin \cite{Paz:07}.

Considering the initial state $\left |\Psi_{\mathcal{SE}}(0)\right
>=(a\left |0\right>+b\left |1\right>)\left |\mathcal{E}(0)\right>$,
where $\left |\mathcal{E}(0)\right>$ is the initial state of the
environment, the evolved reduced density matrix of the system is
\begin{eqnarray}
\rho_\mathcal{S}(t)& = &\mbox{Tr}_{\mathcal{E}}\left
|\Psi_{\mathcal{SE}}(t)\right >\left <\Psi_{\mathcal{SE}}(t)\right |\\\nonumber
& =&|a|^2\left |0\right >\left <0 \right|+ab^*r(t)\left |0\right >\left
<1 \right|\\\nonumber
&+&a^*br^*(t)\left |1\right >\left <0 \right|+|b|^2\left |1\right >\left <1 \right|.
\end{eqnarray}

The off-diagonal terms of the density matrix are modulated by the
decoherence factor $r(t)$ which is the overlap between two states of
the environment obtained by evolving the initial state $\left
  |\Psi(0)\right>$ with two different Hamiltonians,

\begin{equation}
r(t) = \left< \Psi(0) \right| e^{i H_0 t} e^{-i H_1 t} \left| \Psi(0) \right>.
\end{equation}

If the environment is initially in the ground state of $H_{0}$ , $\left|0, g \right>$, the decoherence factor, up
to an irrelevant phase factor, is
\begin{equation}
\label{decohe}
r(t) = \left<0, g \right| e^{-i H_1 t} \left| 0, g \right>.
\end{equation}
This quantity has the same form as the Loschmidt echo or the fidelity, and it contains all the relevant
information about the decoherence process.

To be more specific, let us introduce as an environment a two-level boson system described by a generalized Lipkin
Model, whose Hamiltonian is

\begin{equation}
H_{\mathcal{E}}=\alpha~\widehat{n}_{t}-\frac{1-\alpha}{N}~\widehat{Q}^{\omega}\widehat{Q}^{\omega},
\label{eq:1}
\end{equation}
where the operators $\widehat{n}_{t}$ and $\widehat{Q}^{\omega}$ are defined as

\begin{equation}
\widehat{n}_{t}=t^{\dag}t,\hspace{0.5cm}
\widehat{Q}^{\omega}=s^{\dag}t+t^{\dag}s+\omega~  t^{\dag}t,
\label{eq:2}
\end{equation}
in terms of two species of scalar bosons $s$ and $t$. $\alpha$
and $\omega$ are two independent control parameters, and the total number of
bosons $N=\widehat{n}_{s}+\widehat{n}_{t}$ is a conserved quantity.

It is worth to mention that this two-level bosonic Hamiltonian is
completely equivalent to an SU(2) spin Hamiltonian, with long-range
spin exchange interaction. The equivalence is defined by the 
inverse Schwinger representation of the SU(2) generators
\begin{equation}
S^+=t^{\dagger}s=(S^-)^{\dagger}, \hspace{1cm} S^{z}=\frac{1}{2} (t^{\dagger}t-s^{\dagger}s),
\end{equation}
where $S$ represents the total spin of a chain of $N$ 1/2 spins. 


\subsection{Mean field theory for $H_{\mathcal{E}}$}

In order to study the phase diagram of the Hamiltonian (\ref{eq:1}) as
a function of the control parameters $\alpha$ and $\omega$, it is
usual to rely on a coherent state of the form
\begin{equation}
\left|N, \beta \right> =e^{\sqrt{ \frac {N}{(1+\beta^{2})}}(s^{\dag}+\beta~ t^{\dag})} \left |0\right
>, \label{eq:3}
\end{equation}
where $\left| 0 \right >$ denotes the boson vacuum. The corresponding
energy surface as a function of the variational parameter $\beta$ is
the expectation value of $H_{\mathcal{E}}$ (\ref{eq:1}) in the
coherent state (\ref{eq:3})

\begin{eqnarray}
\label{eq:4}
E(N,\beta)&=&\frac{\langle N,\beta  | H_{\mathcal{E}} | N,\beta \nonumber \rangle}{\langle N,\beta | N,\beta \nonumber \rangle}\\
&=& N \frac{\beta^2}{(1+\beta^2)^2} \Big\{ 5\alpha-4+4 \beta  \omega (\alpha-1)+
\beta^2\big[\alpha+\omega^2(\alpha-1) \big] \Big\} .
\end{eqnarray}

Minimization of the energy (\ref{eq:4}) with respect to $\beta$, for
given values of the control parameters $\alpha$ and $\omega$, gives
the equilibrium value $\beta_{e}$ defining the phase of the system in
the ground state. The value $\beta_{e}=0$ corresponds to the symmetric
phase, and $\beta_{e}\not = 0$ to the broken symmetry
phase.

This Hamiltonian has a second order Quantum Phase Transition (QPT)
along the line $\omega = 0$, and a first order QPT for $\omega \not =
0$. In the latter, the critical point is defined as the situation in
which the minimum in the symmetric phase and in the broken symmetry
phase are degenerate and their energies are equal to zero. The study
of the phase diagram has been done in several publications
\cite{Vidal:06}. Here we summarize its main features.

\begin{itemize}

\item $\beta=0$ is always a stationary point. For $\omega =0$, the
  solution with $\beta=0$ is a maximum for $\alpha<4/5$, and becomes a
  minimum for $\alpha>4/5$. In the case of $\alpha=4/5$, $\beta=0$ is
  an inflection point. $\alpha=4/5$ is the point in which a minimum at
  $\beta=0$ starts to develop and defines the antispinodal line.

 \item For $\omega \neq 0 $ there exists a
  region where two minima, one
  spherical and one deformed, coexist. This region is
  defined by the point where the $\beta=0$ minimum appears
  (antispinodal point) and the point where the $\beta\neq 0$ minimum
  appears (spinodal point). The spinodal line is defined by  the
  implicit equation,
\begin{equation}
\frac{3\,\alpha}{3\,\alpha-4}=\frac{\cal A}{\cal B}
\left (1-\left(1+\frac{\cal B}{\cal A}\right)^{\frac{3}{2}}\right ),
\end{equation}
where ${\cal A}= (4 - 3\,\alpha + 2\,(\alpha -1)\,\omega^2)^2$ and ${\cal B}=36 \,\omega^2 \, ( \alpha -1)^2$. For
example, for $\omega=1/\sqrt{2}$, $\alpha \simeq 0.822559$.

\item In the coexistence region, the critical point is defined by the
  condition that both minima (spherical and deformed) are
  degenerate. At the critical point the two degenerated minima are at
  $\beta_e=0$ and $\beta_e=\omega/2$, and
  their energy is equal to zero. The critical line is therefore defined as
\begin{equation}
\alpha_c=\frac{4+\omega^2}{5+\omega^2}.
\label{eq:5}
\end{equation}
For example, for $\omega=1/\sqrt{2}$, $\alpha_c=9/11$.

\item According to the previous analysis, for $\omega\neq 0$
  there appears a first-order phase transition, while for $\omega=0$
  there is an isolated point of  second-order
  phase transition at $\alpha=4/5$. In this case, antispinodal,
  spinodal and critical points collapse to a single point.

\end{itemize}

\begin{figure}
\begin{center}
\rotatebox{0}{\scalebox{0.55}[0.55]{\includegraphics{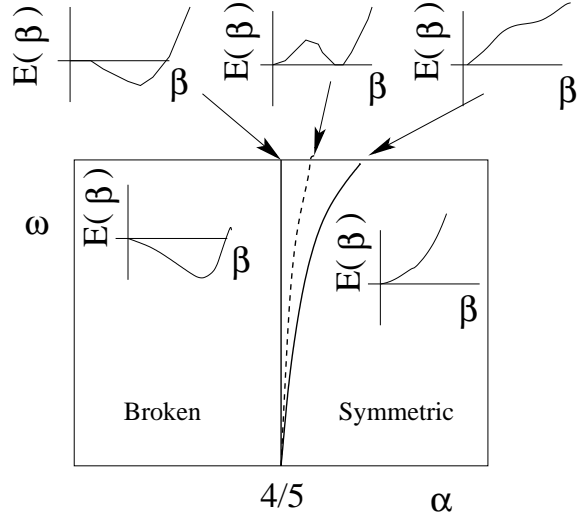}}}
\caption[]{Schematic phase diagram for $H_{\mathcal{E}}$ (\ref{eq:1})
  as a function of the control parameters $\alpha$ and $\omega$.}\label{fig:fases}
\end{center}
\end{figure}

In Fig. \ref{fig:fases} we present a schematic view of the phase
diagram for the environment Hamiltonian $H_{\mathcal{E}}$ (\ref{eq:1})
in the $\omega-\alpha$ plane.

The Hamiltonian (\ref{eq:1}) also displays an Excited State Quantum
Phase Transition (ESQPT), which is analogous to a standard quantum
phase transition, but taking place at some excited critical energy
$E_c$ of the system. We can distinguish between different kinds of
ESQPTs. As it is stated in \cite{Cejnar:07}, in the thermodynamic
limit a crossing of two levels at $E=E_c$ determines a first order
ESQPT, while if the number of interacting levels is locally large at
$E=E_c$ but without real crossings, the ESQPT is continuous. As the
entropy of a quantum system is related to its density of states, a
relationship between an ESQPT and a standard phase transition at a
certain critical temperature can be established in the thermodynamic
limit \cite{Cejnar:09}. These kinds of phase transitions have been
identified in the Lipkin model \cite{Heiss:05}, in the interacting
boson model \cite{Heinze:06}, and in more general boson or fermion
two-level pairing Hamiltonians (for a complete discussion, including a
semiclassical analysis, see \cite{Caprio:08}). In all these cases, the
ESQPT takes place beyond the critical value of the Hamiltonian control
parameter, implying that the critical point moves from the ground
state to an excited state.

\begin{figure}
\begin{center}
\begin{tabular}{cc}
\rotatebox{-90}{\scalebox{0.30}[0.30]{\includegraphics{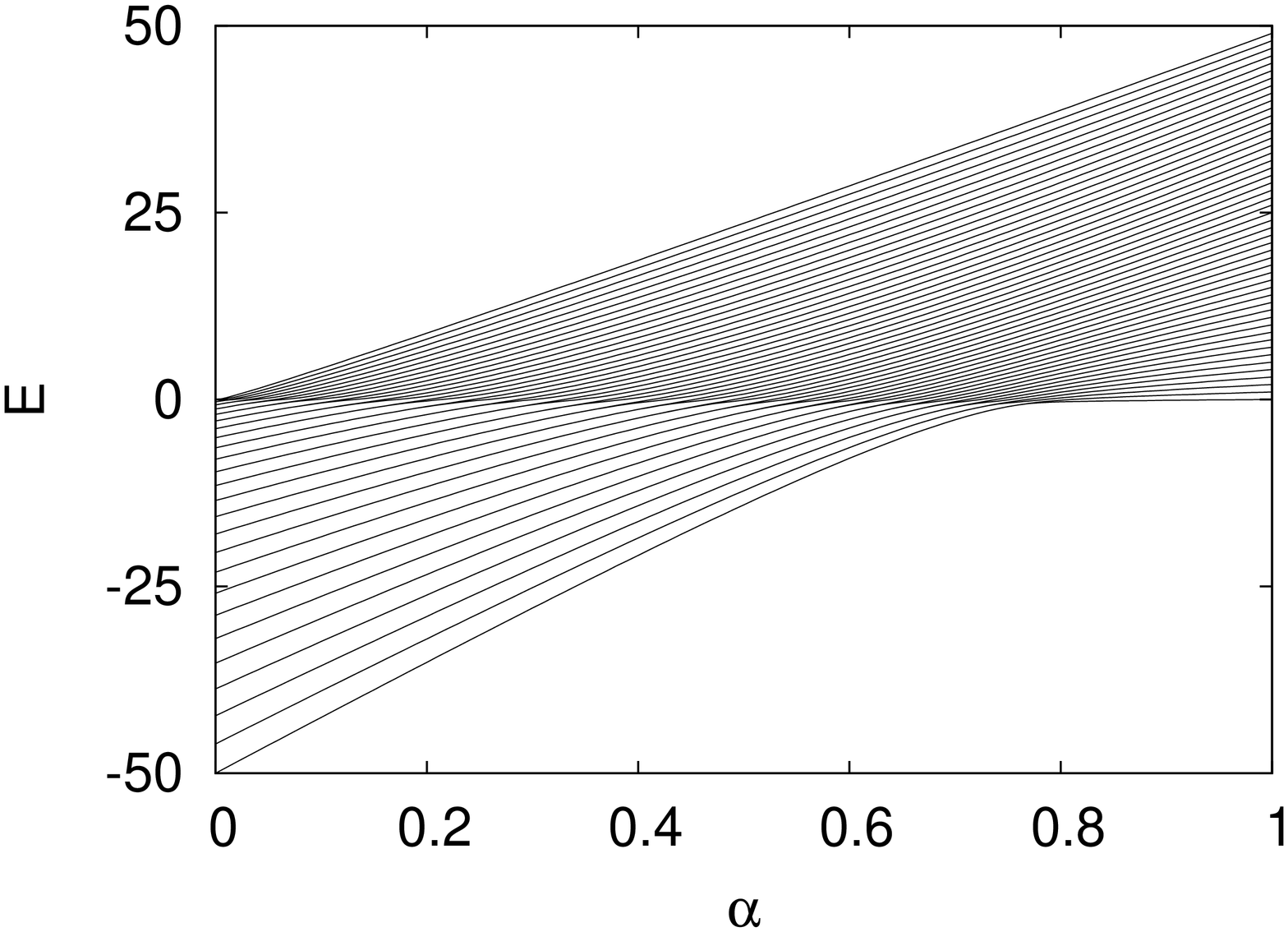}}} &
\rotatebox{-90}{\scalebox{0.30}[0.30]{\includegraphics{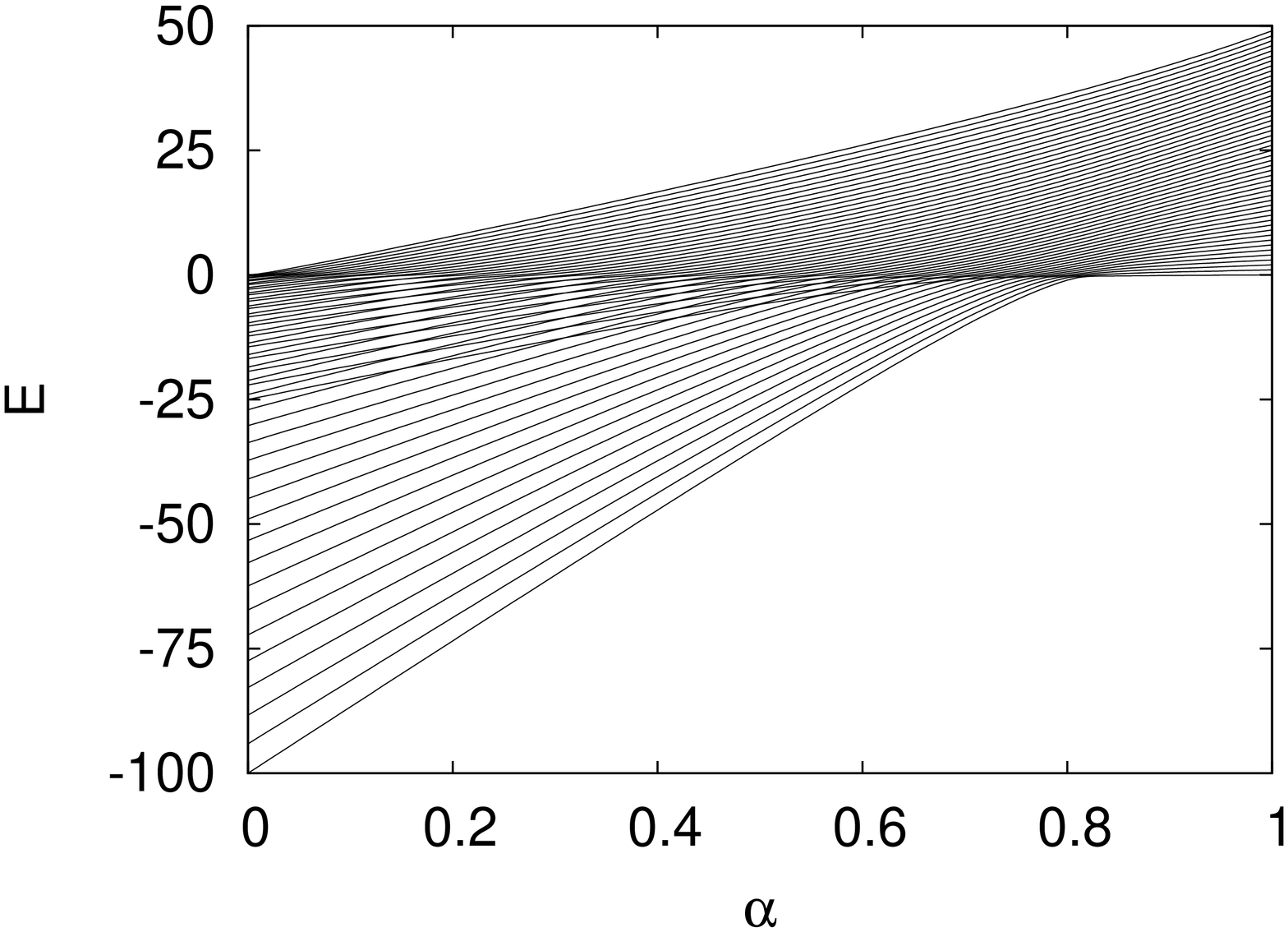}}} \\
(a) $\omega=0$ & (b) $\omega=1/\sqrt{2}$ \\
\end{tabular}
\caption[]{Energy levels of the Hamiltonian (\ref{eq:1}) as a function
  of $\alpha$ for $N=50$ and two different values of $\omega$.}
\label{fig:niveles}
\end{center}
\end{figure}

In Fig. \ref{fig:niveles} we show the energy eigenvalues of the
environmental Hamiltonian (\ref{eq:1}) with $N=50$ bosons as a
function of the control parameter $\alpha$ for $\omega=0$ in left
panel, and $\omega=1/\sqrt{2}$ in right panel. In both cases, we see
for $\alpha<\alpha_c$ a collapse of several levels at $E \approx
0$. In the right panel ($\omega=1/\sqrt{2}$) we can also see a second
critical curve for $E < 0$ that divides the level diagram in two
regions: one in which levels behave smoothly, and another in which the
level density increases and some crossings are observed.

One simple way to analyze the phase diagram is by means of the density
of states. To obtain an analytical approximation for this quantity,
one can start from a coherent state similar to (\ref{eq:3}), in which
real parameter $\beta$ is replaced by the complex parameter
$z=\tan\left(\phi/2\right) \exp\left(i \xi \right)$, in terms of which
the energy is expressed as $\mathcal{H} (\phi,\xi) = \left< N, \phi,
  \xi \right| H \left| N, \phi, \xi \right>$. A good approximation for
the density of states can be obtained by counting how many levels are
there in an energy window $d E$,
\begin{equation}
\rho (E) = \frac{1}{\mathcal{N}} \int d \xi d \phi \left| J
(\phi,\xi) \right| \delta \left( \mathcal{H} (\phi,\xi) - E
\right),
\label{eq:densidad}
\end{equation}
where $\left| J (\phi,\xi) \right|$ is the Jacobian of the
transformation $\left(\phi, \xi \right) \rightarrow (p,q)$, $p$ and
$q$ are the canonical coordinates of the Hamiltonian, and
$\mathcal{N}$ is a normalization constant.

\begin{figure}
\begin{center}
\rotatebox{-90}{\scalebox{0.30}[0.30]{\includegraphics{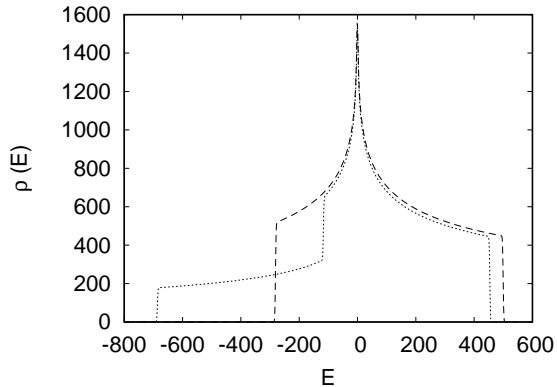}}} \caption[]{Density of states of the
Hamiltonian (\ref{eq:1})
  for $N=1000$, $\alpha=0.5$. The dashed line corresponds to $\omega=0$ and the dotted line to $\omega=1/\sqrt{2}$.}
\label{fig:densidades}
\end{center}
\end{figure}

In Fig. \ref{fig:densidades} we show the density of levels of the
environmental Hamiltonian (\ref{eq:1}), calculated by means of
Eq. (\ref{eq:densidad}), for $N=1000$, $\alpha=1/2$ and the same
values of $\omega$ as in Fig. \ref{fig:niveles}. As it can be seen,
the collapse of levels at $E=0$ gives rise to a cusp singularity of
$\rho(E)$ for both $\omega=0$ and $\omega=1/\sqrt{2}$. In the latter
case, there also exists a jump in the density of states for a fixed
value $E < 0$ ($E\approx -125$ for this value of $\omega$) consistent
with the energy spectra of Fig. \ref{fig:niveles}.  Although not
shown, similar results are obtained for other values of $\alpha$ and
$\omega$. In particular, the jump in the density of states at a
certain value $E<0$ only appears for $\omega >0$. Therefore, two
different kinds of ESQPT exist in excited spectrum of Hamiltonian
(\ref{eq:1}). If we keep the terminology of thermodynamical phase
transitions and we take the number of levels up to an energy E, $N(E)
= \int d E \rho (E)$ as the analogue of the free energy $F(N,T)$, we
can conclude: (a) there exists a continuous $\lambda$ quantum phase
transition at $E^{(2)}_c=0$, for any value of parameter $\omega$; (b)
there also exists a first-order quantum phase transition at some
critical energy $E^{(1)}_c < 0$ if $\omega>0$.

\begin{figure}
\begin{center}
\rotatebox{-90}{\scalebox{0.50}[0.50]{\includegraphics{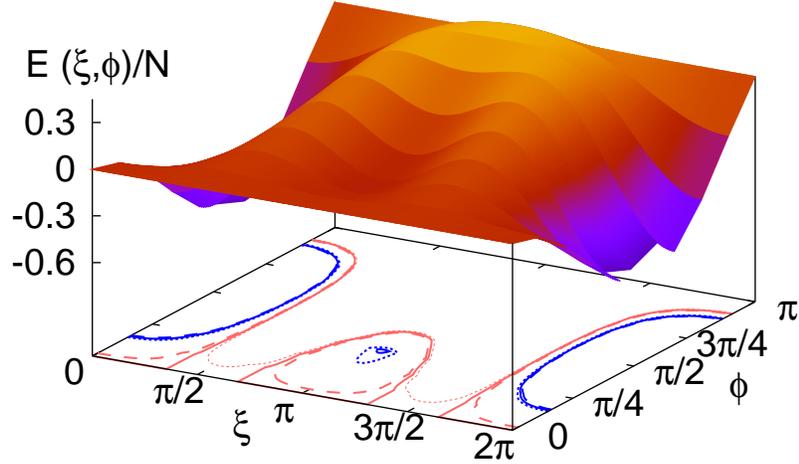}}}
\caption[]{(Color online). Energy surface $\mathcal{H} (\phi,\xi)/N$
  in the thermodynamical limit $N \rightarrow \infty$ for $\alpha=1/2$
  and $\omega=1/\sqrt{2}$. Contour curves are drawn on the base of the
  figure (see text).}
\label{fig:energia}
\end{center}
\end{figure}

To estimate the critical energies at which these quantum phase
transitions take place, we can rely on the energy surface $\mathcal{H}
(\phi,\xi)$. In Fig. \ref{fig:energia} we show $\mathcal{H}
(\phi,\xi)/N$ in the thermodynamical limit $N \rightarrow \infty$ for
$\alpha=1/2$ and $\omega=1/\sqrt{2}$. The curves drawn in the base of
the figure are contour curves for fixed values of the energy
$\mathcal{H} (\phi,\xi)/N = E$. Gray curves (red online) represent
different values of $E$ around $E^{(2)}_c$ for the continuous phase
transition. The solid gray (red online) line represents the critical
point $E_c =0$; this is the only value for which the contour curve is
non-analytic. On the other hand, black curves (blue online) represent
different values of $E$ around the critical energy at which the
first-order ESQPT takes place, $E^{(1)}_c$. In this case, the critical
value is the one at which the island around $\xi=\pi$ appears, that
correspond to a local minimum in the energy surface. This entails
the appearance of another region in the $(\phi,\xi)$ plane
for which the equation $\mathcal{H} (\phi,\xi)/N = E$ has a solution,
and consequently the density of states $\rho(E)$ suddenly increases.

\subsection{Coupling to a single qubit}

Since we are interested in relating the phenomenon of decoherence in a
single qubit with the structure of phases and critical regions in the
environment as defined by the Hamiltonian (\ref{HSE}), we
propose as a coupling Hamiltonian $H_{\lambda_i}=\lambda_{i}
\widehat{n_{t}}$. Choosing $\lambda_{0}=0$ if the qubit is on state
$\left |0 \right >$ and $\lambda_{1}=\lambda$ if the qubit is on the
state $\left |1 \right >$, the effective environment Hamiltonian for
each component of the system results into

\begin{eqnarray}
H_0 &=& \alpha~ \widehat{n}_t - \frac{1-\alpha}{N}~
\widehat{Q}_t\widehat{Q}_t, \label{H0} \\
H_1 &=&(\alpha+\lambda) \widehat{n}_t - \frac{1-\alpha}{N}~ \widehat{Q}_t\widehat{Q}_t~.
\label{H1}
\end{eqnarray}
This means that the qubit only interacts with the environment when it is
on state $\left |1 \right >$.

The system-environment coupling parameter $\lambda$ modifies the
environment Hamiltonian. For certain values of $\alpha$ and $\lambda$,
this modification entails a crossing of the critical lines. Similar
phenomena were previously analyzed by several authors
\cite{Paz:07, Paz:08, Rossini:07}, studying whether a quantum
quench that drives the environment through a QPT implies some kind of
universality in the decoherence process.

Using the coherent state approach \cite{Vidal:06}, it is
straightforward to show that $H_{1}$ goes through a ground state QPT at

\begin{equation}
\lambda_{*}=(1-\alpha)(4+\omega^2)-\alpha
\label{eq:6}
\end{equation}

\noindent for $\alpha < \alpha_{*}$. Therefore, if $\lambda >
\lambda_*$ the quench makes the environment jump from one phase to the
other.

The main purpose of this paper to show that an ESQPT, instead of a
ground state QPT, indeed produces dramatic consequences in the
decoherence process. Using the coherent state approximation, it is
straightforward to obtain that the coupling between the environment
and the qubit entails an energy transfer in the former one, which is
equal to
\begin{equation}
\Delta E (N, \beta, \lambda)= \left< N, \beta \right| \lambda
\widehat{n_t} \left| N, \beta \right> = \lambda N \frac{\beta^2}{1 +
  \beta^2}.
\end{equation}
Therefore, the critical coupling $\lambda_c$ which leads the
environment to the critical energy $E_c$ is
\begin{equation}
E (N, \beta) + \Delta E (N, \beta, \lambda_c) = E_c,
\label{eq:lambdac}
\end{equation}
valid for both first order critical energy $E_c^{(1)}$ and second
order one $E_c^{(2)}$. In general, this is a trascendent equation, and
therefore $\lambda^{(1)}_c$ and $\lambda^{(2)}_c$ (the $\lambda$'s
corresponding to $E_c^{(1)}$ and $E_c^{(2)}$, respectively) have to be obtained
numerically.

\section{Calculation of the decoherence factor}

In order to calculate the decoherence factor (\ref{decohe}) the
expectation value of $H_1$ (\ref{H1}) in the ground state $|0, g
\rangle$ of $H_0$ (\ref{H0}) is needed. The decoupling of the complete
system-environment Hamiltonian into the independent Hamiltonians $H_0$
and $H_1$ for each qubit state allows an exact diagonalization for
large systems.  In the following two subsections we will describe the
exact formalism and make a comparison with mean field techniques
supplemented with a Tamm-Dankoff approximation (TDA) treatment of the
excited spectrum.

\subsection{Exact diagonalization}

A general Hamiltonian in terms of $s$ and $t$ bosons including up to
two body terms is,
\begin{eqnarray}
\nonumber
H_{st}&=& a t^\dag t + b(t^\dag s +s^\dag t)+ c t^\dag s s^\dag t\\
&+& d (t^\dag s t^\dag s +s^\dag t s^\dag t)
+e (t^\dag s t^\dag t+ t^\dag t s^\dag t)+ f t^\dag t t^\dag t
\label{lip-ham1}
\end{eqnarray}
where $a,b,c,d,e$ and $f$ are arbitrary parameters.


Both Hamiltonians, $H_0$ (\ref{H0}) and  $H_1$ (\ref{H1}), are
particular cases of $H_{st}$ (\ref{lip-ham1}) with the
following parameters,
\begin{eqnarray}
a&=&\alpha+\lambda -2\frac{\alpha-1}{N} \nonumber \\
b&=&\omega\frac{\alpha-1}{N} \nonumber \\
c&=&2\frac{\alpha-1}{N} \nonumber \\
d&=&\frac{\alpha-1}{N} \nonumber \\
e&=&2 \omega \frac{\alpha-1}{N} \nonumber \\
f&=& \omega^2 \frac{\alpha-1}{N} \nonumber \\
\Delta&=& \alpha-1,
\end{eqnarray}
where $\Delta$ is an irrelevant global shift in energy.

The exact diagonalization of the $st$ Hamiltonian (\ref{lip-ham1}), and consequently of $H_0$ and $H_1$, reduces
to the diagonalization of a tridiagonal matrix in the basis
\begin{equation}
|N l \rangle =\frac{ {t^\dag}^l {s^\dag}^{N-l}}{\sqrt{l!
    (N-l)!}} |0\rangle,
\end{equation}
where $|0\rangle $ is the boson vacuum and $0\leq l \leq N$. Therefore, the dimension of the Hamiltonian matrix is
$d=N+1$.

 The relevant matrix elements are,
\begin{eqnarray}
\langle N l | H_{st} | N l \rangle &=& a l + f l^2 +c l (1+N-l),\\
\langle N l | H_{st} | N l+1 \rangle &=&b \sqrt{(N-l)(l +1)}+e \sqrt{(l+1)(N-l)}~l,\\
\langle N l | H_{st} | N l+2 \rangle &=&d\sqrt{(l+2)(l+1)}\sqrt{(N-l)(N-l-1)},
\end{eqnarray}
being all the others equal to zero. The diagonalization of the
corresponding tridiagonal matrix can be done easily even for large $N$
values, providing the exact results for the eigenenergies and
eigenfunctions of $H_0$ and $H_1$ and consequently allowing to
calculate numerically $r(t)$.

\subsection{The Tamm-Dankoff approximation}

Before applying the exact diagonalization techniques to study the
behavior of the decoherence as a fuction of the set of model
parameters and particularly in relation to the quantum phase
transitions (first and second order) in the ground (QPT) and excited
states (ESQPT) of the environment, we will introduce an extension of
the mean field approximation based on the TDA but including two phonon
anharmonicities.

Let us consider the condensate boson of the state \eqref{eq:3} as a
ground state deformed boson in a rotated basis.  Since two
Hamiltonians are involved, $H_0$ and $H_1$, let us formulate the
approximation for both in terms of a generic $H_i$ ($i=0,1$). The
variational parameter $\beta$ in the condensate could be different for
both Hamiltonians; therefore, the notation $\beta_{i}$ ($i=0,1$) will
be used to distinguish between both cases. With this notation, the
deformed bosons ($g$ and $e$) for $H_i$ are related to the initial
ones ($s$ and $t$ bosons) by

\begin{eqnarray}
\label{eq:7}
\Gamma ^{\dag}_{i,g}&=&\frac{1}{\sqrt{1+\beta_{i}^2}}(s^{\dag}+\beta_{i} t^{\dag}),\\
\Gamma ^{\dag}_{i,e}&=&\frac{1}{\sqrt{1+\beta_{i}^2}}(-\beta_{i} s^{\dag}+ t^{\dag}).
\end{eqnarray}

In terms of the deformed bosons the ground state and the first excited states are
\begin{eqnarray}
\label{eq:8}
\left |i, g \right
>&=&\frac{1}{\sqrt{N!}}\left(\Gamma_{i,g}^{\dag}\right)^N \left
|0\right > , \\
\left |i, e \right
>&=&\frac{1}{\sqrt{(N-1)!}}\Gamma_{i,e}^{\dag}(\Gamma_{i,g}^{\dag})^{N-1}\left
|0\right >.
\end{eqnarray}

In this framework, higher excited states can be constructed by
directly replacing a ground state boson condensate by an excited
$\beta$ boson; this procedure is known as the Tamm-Dancoff
approximation (TDA) method. In addition, with this basis is possible
to write a diagonal Hamiltonian in terms of the new bosons. If only
one body terms are included,

\begin{eqnarray}
\label{eq:9} H_i&\approx &\left<i,g|H_i|i, g\right >+ \Big ( \left<i, e|H_i|i, e\right >-\left <i,g|H_i|i,g\right
> \Big )\Gamma^{\dag}_{i,e}\Gamma_{i,e}\\
& &=E_{i,0}+\Delta_{e_i}\Gamma^{\dag}_{i,e}\Gamma_{i,e},\nonumber
\end{eqnarray}
where $E_{i,0}=\left<i,g|H_i|i,g\right >$ and $\Delta_{e_i}=\left (\left <i,e|H_i|i,e\right >-\left
<i,g|H_i|i,g\right > \right )$.

The calculation for $r(t)$ \eqref{decohe} involves the $H_0$ ground
state and the Hamiltonian $H_1$. Thus, it is necessary to relate the
intrinsic bosons for $H_0$ and $H_1$. The relation between both boson
families is given by

\begin{eqnarray}
\label{eq:11}
\Gamma^{\dag}_{0,g}&=&\sum_{p} ~f_{gp} ~ \Gamma_{1,p}^{\dag} , \\
\Gamma^{\dag}_{0,e}&=&\sum_{p} ~f_{\beta p} ~ \Gamma_{1,p}^{\dag} ,
\end{eqnarray}

\noindent where this sum is for $p=g$ and $p=e$, and the
coefficients of the needed transformation are,

\begin{eqnarray}
\label{eq:12}
f_{gg}&=&\frac{1}{\sqrt{1+\beta_{0}^{2}}}\frac{1}{\sqrt{1+\beta_{1}^{2}}}(1+\beta_{0}\beta_{1}),\\
f_{ge}&=&\frac{1}{\sqrt{1+\beta_{0}^{2}}}\frac{1}{\sqrt{1+\beta_{1}^{2}}}(\beta_{0}-\beta_{1}).
\end{eqnarray}

With the preceding transformation it is possible to write $\left
|0, g\right >$ in terms of the $H_1$ intrinsic bosons

\begin{equation}
\label{eq:13}
\left |0, g\right
>=\frac{1}{\sqrt{N!}}\left(f_{gg}\Gamma_{1,g}^{\dag}+
f_{ge}\Gamma_{1,e}^{\dag} \right)^{N}\left |0\right >.
\end{equation}

Using the binomial expansion of \eqref{eq:13} is then straightforward to calculate the decoherence factor $r(t)$
using the TDA basis up to an irrelevant phase factor

\begin{equation}
\label{eq:14}
r(t)=\sum_{k=0}^{N}\binom{N}{k} ~ \left(f_{gg}\right)^{2(N-k)}~
  \left(f_{ge}\right)^{2k} ~e^{-i\Delta_{e_1}
  kt},
\end{equation}
where $\Delta_{e_1}=\left (\left <1,e|H_1|1,e\right >-\left <1,g|H_1|1,g\right > \right )$. A more compact
expression for $r(t)$ can be obtained using the transformation,

\begin{equation}
\label{eq:15} e^{-i\Delta_{e_{1}}tk}=e^{-i(\Delta_{e_{1}}/2)tN} ~ e^{i(\Delta_{e_{1}}/2)t(N-k)} ~
e^{-i(\Delta_{e_{1}}/2)tk}.
\end{equation}

Therefore, the decoherence factor $r(t)$ in the TDA reduces to

\begin{equation}
\label{eq:16} r(t)=e^{-i(\Delta_{e_{1}}/2)tN}\left(\left(f_{gg}\right)^{2} ~ e^{i(\Delta_{e_{1}}/2)t}+
\left(f_{ge}\right)^{2} ~ e^{-i(\Delta_{e_{1}}/2)t}\right)^{N}.
\end{equation}

The matrix elements required for calculating $r(t)$ are

\begin{eqnarray}
\nonumber
\langle i,g |H_i |i,g \rangle &=& (a+c+f)\frac{\beta_i^2}{1+\beta_i^2}
N +2 b \frac{\beta_i}{1+\beta_i^2} N \\
&+& \frac{N(N-1)}{(1+\beta_i^2)^2} \Big (
(c+2d)\beta_i^2+2 e \beta_i^3 +f \beta_i^4
\Big ),
\label{Delta1}
\end{eqnarray}
and
\begin{eqnarray}
\nonumber \langle i, e |H_i |i, e \rangle&=& \frac{1}{1+\beta_i^2} (a+c+f-2
b\beta_i)+c \frac{(1-\beta_i^2)^2}{(1+\beta_i^2)^2} (N-1)\\
\nonumber
&+& \frac{4(N-1)}{(1+\beta_i^2)^2} (f \beta_i^2-2 d\beta_i^2+e
(\beta_i-\beta_i^3))\\
\nonumber
&+&(a+c+f)\frac{\beta_i^2}{1+\beta_i^2}
(N-1) +2 b \frac{\beta_i}{1+\beta_i^2} (N-1) \\
&+& \frac{(N-1)(N-2)}{(1+\beta_i^2)^2} \Big (
(c+2d)\beta_i^2+2 e \beta_i^3 +f \beta_i^4
\Big ) .
\label{Delta1b}
\end{eqnarray}

A simple inspection reveals that decoherence factor $r(t)$
(\ref{eq:16}) does not give a good approximation of the exact results
(see below and \cite{Relano:08}). The modulus of $r(t)$ is
\begin{equation}
\left| r(t) \right| = \left| \left( f_{gg}^2 \right) e^{i \left(
  \Delta_{e_1}/2 \right) t} + \left( f_{ge}^2 \right) e^{-i
  \left( \Delta_{e_1}/2 \right) t} \right|^N = \left| f_{ge}
\right|^{2N} \left| \left( \frac{f_{gg}}{f_{ge}} \right)^2 + e^{i
  \Delta_{e_1} t} \right|^N.
\end{equation}
As a particular example, let us consider $\beta_1 = 0$ and $\beta_0
\neq 0$, that is, the situation in which the coupling of the qubit to
the environment forces the environments to cross the phase transition
from the broken phase to the symmetric phase. In this situation
\begin{eqnarray}
f_{gg} &=& \frac{1}{\sqrt{1 + \beta_0^2}} \\
f_{ge} &=& \frac{\beta_0}{\sqrt{1 + \beta_0^2}}.
\end{eqnarray}
From these expressions, it is straightforward to obtain that $\left| r(t) \right|$ oscillates between
\begin{eqnarray}
\left| r(t) \right|_{\text{max}} &=& 1; \\
\left| r(t) \right|_{\text{min}}&=& \left| \frac{\beta_0^2 -1}{\beta_0^2+1} \right|^N \longrightarrow 0, \; \text{for} \; N \longrightarrow \infty.
\end{eqnarray}

Therefore, we can conclude that TDA approximation including just one
phonon excitations does not account for the decay of the envelope of
the decoherence factor reported in \cite{Relano:08} (see below for
more details). This evidence suggests to go further within the spirit
of TDA by including the anharmonicities of the two-phonon
excitations. For this purpose it is needed to construct the states two
TDA excitations as

\begin{eqnarray}
\label{eq:17} \left |i,e^{2}\right
>&=&\frac{1}{\sqrt{2}}\frac{1}{\sqrt{(N-2)!}}(\Gamma_{i,e}^{\dag})^2(\Gamma_{i,g}^{\dag})^{N-2}\left
|0\right >.
\end{eqnarray}

From this state we derive the diagonal part of the Hamiltonian as

\begin{equation}
\begin{split}
\label{eq:18} H_i &\approx \element{i,g}{H_i}{i,g}+\Big (\element{i,e}{H_i}{i,e}-\element{i,g}{H_i}{i,g} \Big
)\Gamma^{\dag}_{i,e}\Gamma_{i,e}\\
& +\left ( \frac{\element{i, e
    ^{2}}{H_i}{i,e ^{2}}}{2}-\element{i,e}{H_i}{i,e}+\frac{\element{i,g}{H_i}{i,g}}{2} \right ) \Gamma^{\dag}_{i,e}\Gamma^{\dag}_{i,e}\Gamma_{i,e}\Gamma_{i,e}\\
& =E_{i,0}+\Delta_{e_i}\Gamma^{\dag}_{i,e}\Gamma_{i,e}+\Omega
_{e_i}\Gamma^{\dag}_{i,e}\Gamma^{\dag}_{i,e}\Gamma_{i,e}\Gamma_{i,e},
\end{split}
\end{equation}

\noindent where $E_{i,0}=\element{i,g}{H_i}{i,g}$, $\Delta_{e_i}=\left
(\element{i,e}{H_i}{i,e}-\element{i,g}{H_i}{i,g} \right )$, and $\Omega _{e_i}=\left (
\frac{\element{i,e^{2}}{H_i}{i,e^{2}}}{2}-\element{i,e}{H_i}{i,e}+\frac{\element{i,g}{H_i}{i,g}}{2} \right )$.

With the preceding transformation (\ref{eq:13}) we can obtain the decoherence factor in the improved approximation
up to an irrelevant phase factor

\begin{equation}
\label{eq:19} r(t)=\sum_{k=0}^{N}\binom{N}{k}~ \left(f_{gg}\right)^{2(N-k)} ~ \left(f_{ge}\right)^{2k} ~
e^{-i(\Delta_{e_1}k+\Omega_{e_1}k(k-1))t}.
\end{equation}

In addition to (\ref{Delta1}) and  (\ref{Delta1b}), the only needed matrix element for obtaining $r(t)$ is
$\Omega_{e_1}$ which follows from $\langle i,g |H_i |i,g \rangle$, $\langle i, e |H_i |i, e \rangle$, given above
and

\begin{eqnarray}
\nonumber \langle i,e^2 |H_i |i, e^2 \rangle&=& \frac{1}{(1+\beta_i^2)^2}\Big( (2 c+ 4 d) \beta^2_i +4 e \beta_i^3
+2 f \beta_i^4 \Big )\\
\nonumber
&+&\frac{2}{1+\beta_i^2} (a+c+f-2
b\beta_i)+2 c \frac{(1-\beta_i^2)^2}{(1+\beta_i^2)^2} (N-2)\\
\nonumber
&+& \frac{8(N-2)}{(1+\beta_i^2)^2} (f \beta_i^2-2 d\beta_i^2+e
(\beta_i-\beta_i^3))\\
\nonumber
&+&(a+c+f)\frac{\beta_i^2}{1+\beta_i^2}
(N-2) +2 b \frac{\beta_i}{1+\beta_i^2} (N-2) \\
&+& \frac{(N-2)(N-3)}{(1+\beta_i^2)^2} \Big (
(c+2d)\beta_i^2+2 e \beta_i^3 +f \beta_i^4
\Big ) .
\label{Omega}
\end{eqnarray}

Inserting (\ref{Delta1}), (\ref{Delta1b}) and (\ref{Omega}) into
(\ref{eq:19}) we arrive to the final form of the decoherence factor
$r(t)$ within the extended TDA approximation. In this case, a
semi-quantitative analysis as the previous one cannot be easily
done. A comparison with exact numerical calculations is performed in next
section.

\section{Results}

In this section we present the main features of evolution of the
system described by (\ref{HSE}) under the influence of the environment
given by (\ref{eq:1}). A brief report of the relationship between the
decoherence in the qubit and the excited state quantum phase
transitions in the environment was given in \cite{Relano:08}. Here, we
extend the analysis, comparing the numerical results with the
Tamm-Dankoff approximation, and also facing the case of $\omega \neq
0$, that was not considered in \cite{Relano:08}. As two paradigmatic
cases, we deal with the cases $\alpha=1/2$, and $\omega=0$ and
$\omega=1/\sqrt{2}$. Different choices for the defining parameters of
the model give rise to the same qualitative results.

\subsection{Decoherence factor for the Continuous ESQPT}

All the information about the decoherence process induced by the
environment (\ref{eq:1}) in the central qubit is encoded in the
decoherence factor (\ref{decohe}). As mentioned above, for the
Hamiltonian we are using there is always a continuous ESQPT
independently of the value of $\omega$. In addition, for $\omega \neq
0$ there also appears a first order ESQPT. In this subsection we will
analyze the effect of the continuous ESQPT on the decoherence factor,
while the effect of the first order ESQPT on the decoherence factor
will be discussed in the next subsection.

\begin{figure}
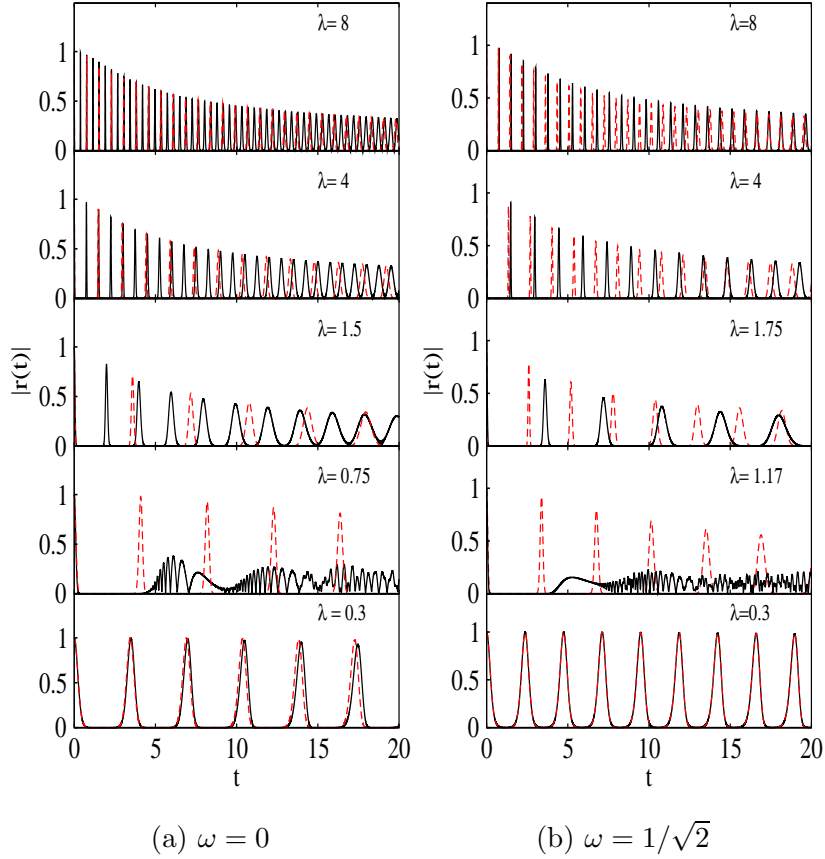

\begin{center}
\begin{tabular}{cc}
{\scalebox{0.50}[0.65]{\includegraphics{figure5a.eps}}} &
{\scalebox{0.50}[0.65]{\includegraphics{figure5b.eps}}} \\
(a) $\omega=0$ & (b) $\omega=1/\sqrt{2}$ \\
\end{tabular}
\caption[]{(Color online)
$|r(t)|$ for $\alpha=1/2$ and five
different values of $\lambda$ for two selections of the coupling
system-environment parameter $\omega$, $\omega=0$ on the left and
$\omega=1/\sqrt{2}$ on the right. In all cases $N=1000$. Solid
(black) lines correspond to the exact solution, and dashed (red) lines
to the TDA calculation.} \label{fig:r(t)}
\end{center}
\end{figure}

In Fig. \ref{fig:r(t)} we show the results for the decoherence factor for $N=1000$ bosons and
$\alpha=1/2$, for two $\omega$ values, $\omega=0$ (left panels) and $\omega=1/\sqrt{2}$ (rigth
panels). The objective of this figure is to show the effect of the
continuous ESQPT on the decoherence factor. Solving Eq. (\ref{eq:lambdac}) for
the value of $\alpha=1/2$, the continuous ESQPT ($E_c^{(2)}=0$)
takes place at $\lambda_c^{(2)}=0.75$ for $\omega=0$ (left panel) and at
$\lambda_c^{(2)}=1.17$ for $\omega=1/\sqrt{2}$ (right panel).
Several features deserve to be discussed. First of all, we can
see that the TDA calculation works pretty well for small and large
values of $\lambda$. In particular, the shape of the envelope, which
remains unaffected by the increase of $\lambda$ for $\lambda \gg
\lambda_c^{(2)}$, is very well described by the TDA calculation (see panels
for $\lambda=4$ and $\lambda=8$ in Fig. \ref{fig:r(t)}). Since this
approximation mainly relies on the position of the first and the
second excited states of $H_{\mathcal{E}}$, we can conclude that the
information contained in the low energy spectrum is enough to have a good
idea about the properties of the highest excited levels of the
environmental Hamiltonian. Note that switching on the interaction
between the central qubit and the environment entails an effective
increase of the environmental energy roughly given by $\Delta E =
\left< g_0 \right| H_1 (\lambda) \left| g_0 \right> - E_0$, and
therefore a large value of $\lambda$ implies that the state of the
environment {\it jumps} from the ground state to a mixed high-energy
state.

On the other hand, as it is clearly shown in the left panels
corresponding to $\lambda=\lambda_c^{(2)}=0.75$ and
$\lambda=\lambda_*=1.5$, and the right panels
$\lambda=\lambda_c^{(2)}=1.17$ and $\lambda=\lambda_*=1.75$,
the TDA calculations fails
for intermediate values of $\lambda$. These two values correspond to
the critical couplings $\lambda_c^{(2)}$ and $\lambda_*$, corresponding to
the excited state and the ground state quantum phase transitions,
given by Eqs. (\ref{eq:lambdac}) and (\ref{eq:6}) respectively. The
reason why the Tamm-Dankoff approximation does not work for these
values is straightforward. The ESQPT entails a singularity in the
energy spectrum far above the first excited state, which gives rise to
the main contribution in the TDA calculation. On the other hand, the
ground state QPT does not affect the decoherence suffered by the
central qubit because the coupling $\lambda$ makes the environment to
jump far above the critical point which entails a singularity in
the gap between the ground and the first excited states. However, as
the TDA calculation for $r(t)$ strongly depends on this gap, it is
spuriously affected by the QPT induced by the critical coupling
$\lambda_*$.

Finally, the best agreement between the Tamm-Dankoff approximation and
the exact calculation happens for $\lambda=0.3$, far below
$\lambda_c^{(2)}$. Not only the envelope of the decoherence factor is
well reproduced, but also the positions of the local maximum are well
placed. In this case, the small coupling makes the environment to jump
from the ground state to a mixed low-energy state. Therefore, it is
reasonable to assume that the description provided by the TDA, which
only takes in consideration the first excited state and a global
measure of the anharmonicites of the spectrum, is a better
approximation for small values of $\lambda$.

\subsubsection{Analysis of the critical behavior of the decoherence at
  the continuous ESQPT}

As it is shown in Fig. \ref{fig:r(t)}, the decoherence of the central
qubit behaves in a singular way for a critical coupling $\lambda_c^{(2)}$,
which makes the environment to jump to the critical energy
$E^{(2)}_c=0$. As the density of states in both cases $\omega=0$ and
$\omega \neq 0$ display the same critical behavior around this value
(see Fig. \ref{fig:densidades}), also the same singular behavior for
the decoherence is expected.

\begin{figure}
\begin{center}
\begin{tabular}{cc}
\rotatebox{-90}{\scalebox{0.30}[0.30]{\includegraphics{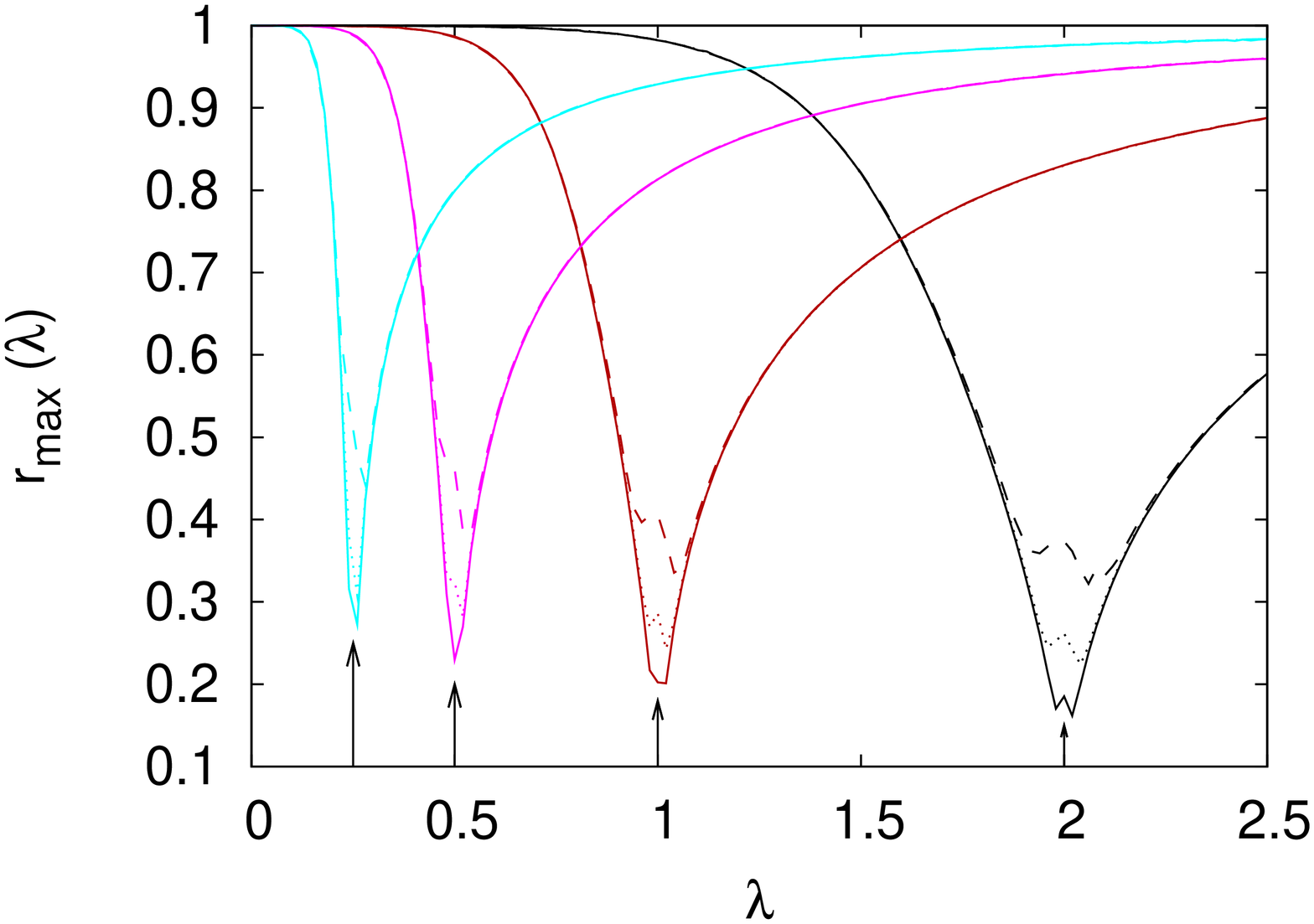}}} &
\rotatebox{-90}{\scalebox{0.30}[0.30]{\includegraphics{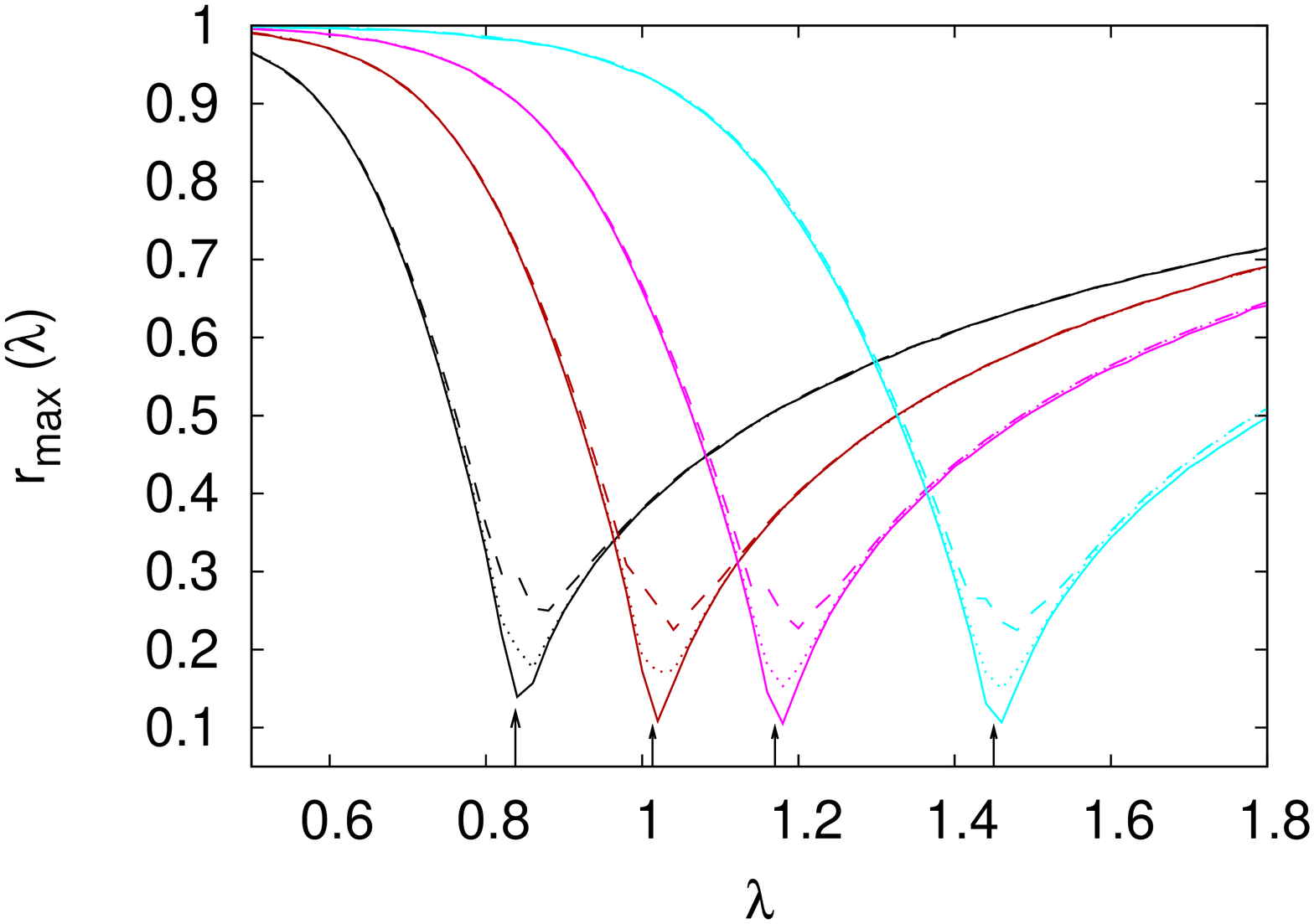}}} \\
(a) $\omega=0$ & (b) $\alpha=1/2$ \\
\end{tabular}
\caption[]{(Color online) $r_{max}$ in function of the coupling $\lambda$, for
different values of $\alpha$, $\omega$, and N. In left panel $\omega=0$. Black lines represent the case $\alpha=0$;
dark gray (red online) lines, $\alpha=0.4$; gray (magenta online), $\alpha=0.6$;
and light gray (cyan online), $\alpha=0.7$. Solid lines represent N
=10 000; dotted lines, N=2500; and dashed lines, N=600. In right
panel, $\alpha=1/2$, and lines represent the cases $\omega=0.2$,
$\omega=1/2$, $\omega=1/\sqrt{2}$, and $\omega=1$, with the same color
code than right panel. Arrows show the critical coupling $\lambda_c$
provided by Eq. (\ref{eq:lambdac}).}
\label{fig:rmax}
\end{center}
\end{figure}

In. Fig. \ref{fig:rmax} we show $r_{max} (\lambda)$, defined as the
second maximum of $| r(t) |$ (the first maximum is trivially
$|r(t=0)|=1$).  The left panel displays the case $\omega=0$ for several
values of $\alpha$, and the right panel the case $\alpha=1/2$, for
several values of $\omega \neq 0$ (see caption for details). We can
see that the behavior of this quantity is the same for $\omega=0$ and
$\omega \neq 0$. It evolves smoothly and independently of the size of
the system $N$ for values far from the critical coupling
$\lambda_c^{(2)}$, provided by Eq. (\ref{eq:lambdac}) and shown in
Tab. \ref{tab:lambdac}. In a small region around $\lambda \sim
\lambda_c^{(2)}$, $r_{max}$ becomes sharp, and the value of the minimum
depends on the size of the system $N$; the larger is the system, the
smaller is $r_{max} (\lambda_c^{(2)})$. Therefore, for both $\omega=0$
and $\omega \neq 0$, the decoherence factor behaves in a critical way
around $\lambda=\lambda_c^{(2)}$ where $r_{max}(\lambda_c^{(2)})$
undergoes a dip towards zero which is sharper and deeper for larger
values of N.

\begin{table}
\begin{tabular}{cccc|cccc}\hline\hline
\multicolumn{4}{c|}{$\omega=0$} &
 \multicolumn{4}{|c}{$\alpha=1/2$} \\ \hline
$\alpha=0$ & $\alpha=0.4$ & $\alpha=0.6$ & $\alpha=0.7$ & $\omega=0.2$
 & $\omega=0.5$ & $\omega=1/\sqrt{2}$ & $\omega=1$ \\
 $\lambda_c^{(2)}=2$ & $\lambda_c^{(2)}=1$ & $\lambda_c^{(2)}=0.5$ & $\lambda_c^{(2)}=0.25$ &
 $\lambda_c^{(2)}=0.83$ & $\lambda_c^{(2)}=1.01$ & $\lambda_c^{(2)}=1.17$ &
 $\lambda_c^{(2)}=1.45$ \\
    \hline \hline \\
\end{tabular}
\caption[]{Critical couplings $\lambda_c^{(2)}$ for the eight cases depicted in Fig. \ref{fig:rmax}}
\label{tab:lambdac}
\end{table}

We now investigate the thermodynamical limit, by performing a finite
size scaling analysis. The largest system that we could treat exactly
has a size of around $N=10000$; going beyond this value is very
difficult since for a complete calculation of $r_{max}
(\lambda_c^{(2)})$ all the eigenvalues and eigenvectors of the
environmental Hamiltonian are needed. Starting with systems of
$N=100$, we analize the finite size scaling along two orders of
magnitude.

\begin{figure}
\begin{center}
\begin{tabular}{cc}
\rotatebox{-90}{\scalebox{0.30}[0.30]{\includegraphics{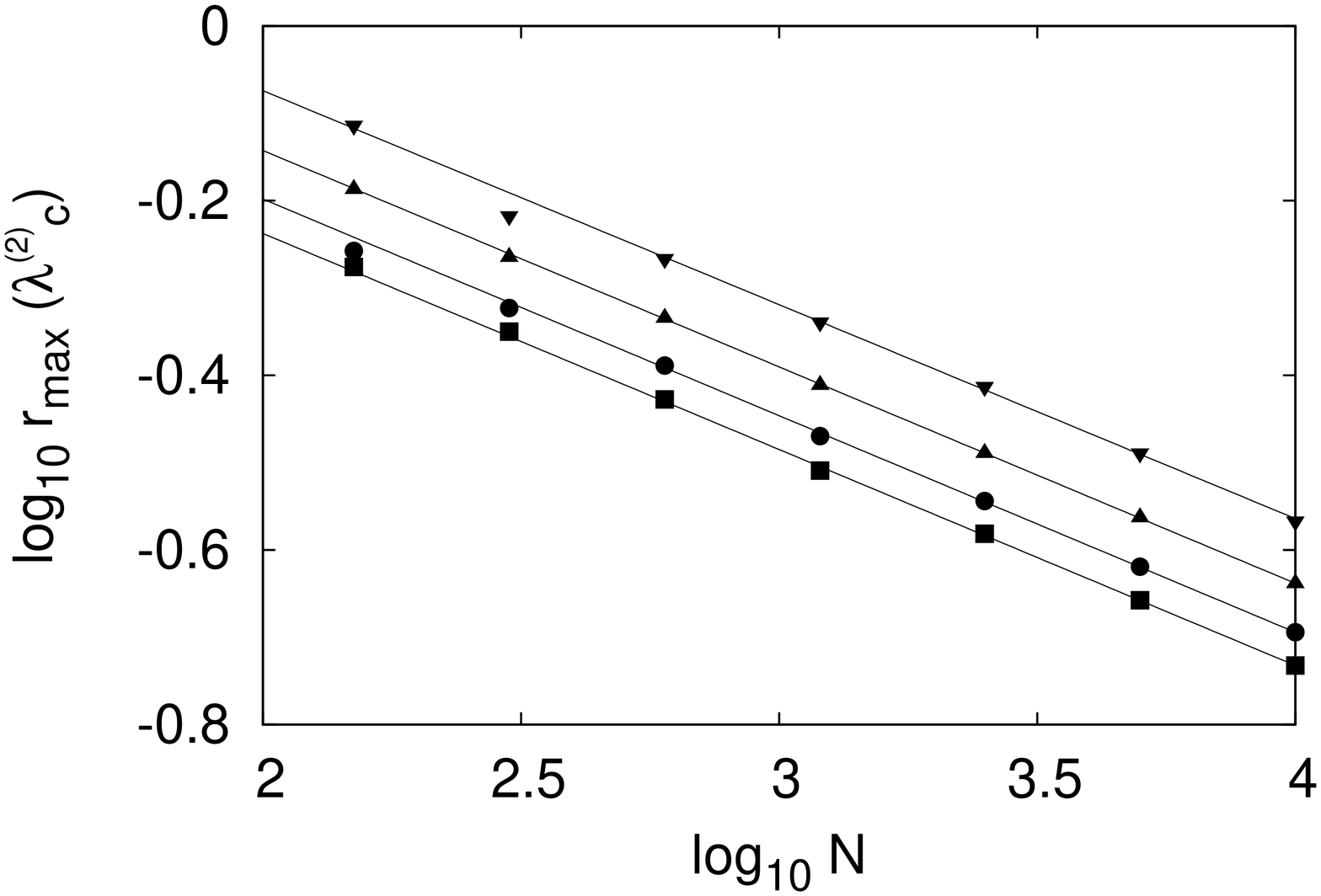}}} &
\rotatebox{-90}{\scalebox{0.30}[0.30]{\includegraphics{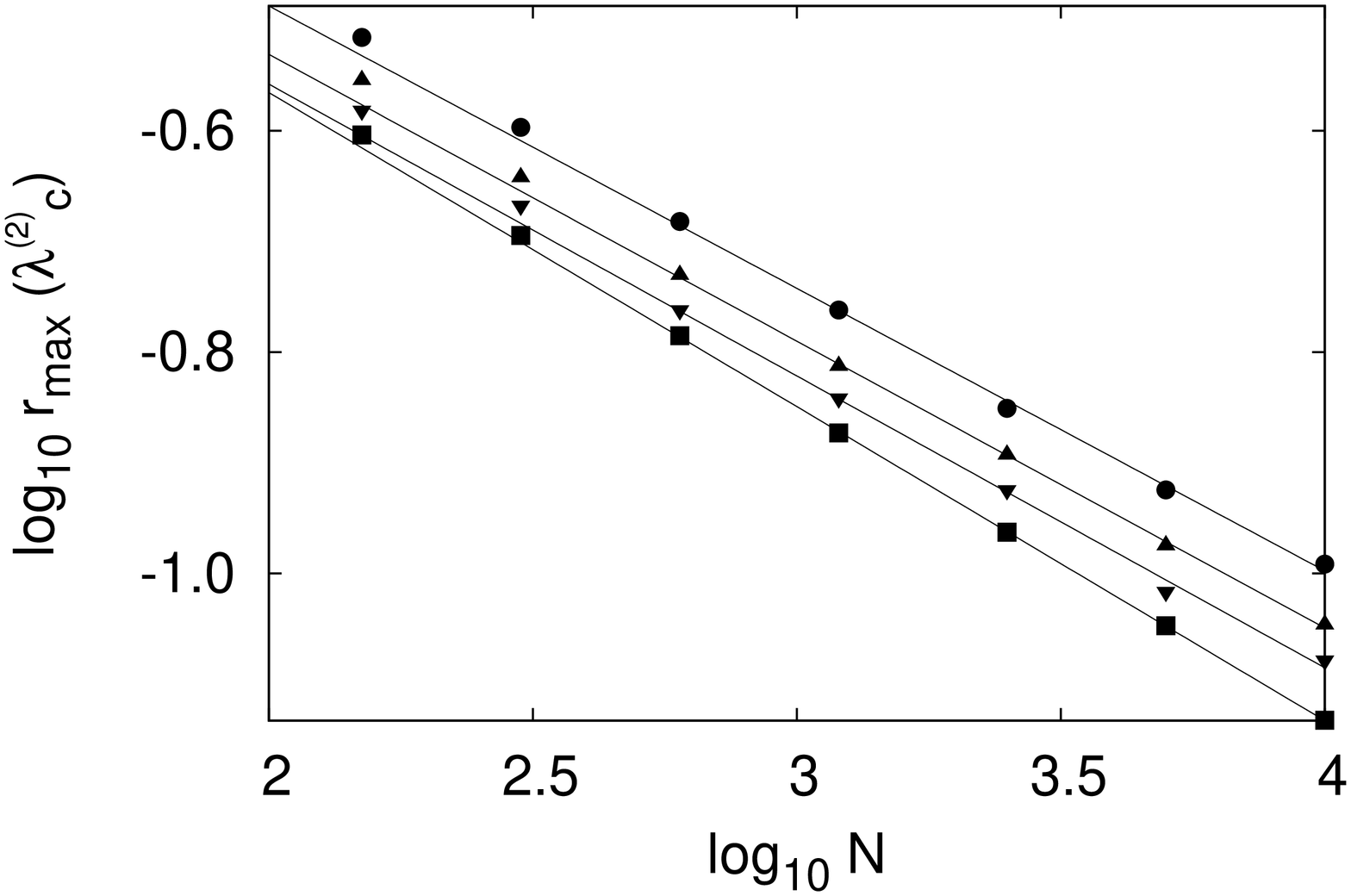}}} \\
(a) $\omega=0$ & (b) $\alpha=1/2$ \\
\end{tabular}
\caption[]{$r_{max}(\lambda_c^{(2)})$ in function of the size of the environment $N$, in
a double logarithmic scale. Left panel represents $\omega=0$; right
panel $\alpha=1/2$. Squares represent the case $\alpha=0$ (left) and
$\omega=0.2$ (right); circles,
$\alpha =0.4$ (left) and $\omega=0.5$ (right); upper triangles,
$\alpha =0.6$ (left) and $\omega=1/\sqrt{2}$ (right); lower triangles,
$\alpha =0.7$ (left) and $\omega=1$ (right). Straight lines represent
the best fit to a power law $r_{max} (\lambda_c^{(2)}) = A N^{-\gamma}$.}
\label{fig:exponentes}
\end{center}
\end{figure}

In Fig. \ref{fig:exponentes} we show how $r_{max} (\lambda_c^{(2)})$
evolves with the size $N$ of the environment, both for $\omega=0$ and
several values of $\alpha$ (left panel), and $\alpha=1/2$ and several
values of $\omega \neq 0$. In all the cases, a power-law scaling
$r_{\max} (\lambda_c^{(2)}) \sim N^{-\gamma}$ is observed, and
therefore we can expect that $r_{max} (\lambda_c^{(2)}) \rightarrow 0$
in the thermodynamical limit $N \rightarrow \infty$. Nevertheless,
subtle differences between verying $\alpha$ with $\omega=0$ and
varying $\omega$ with $\alpha=1/2$ are observed. The results for the
exponent $\gamma$, shown in Tab. \ref{tab:exponentes}, are very close
to the proposed $\gamma = 1/4$ \cite{Relano:08} for
$\omega=0$. However, the numerical estimates seem to increase for
larger values of $\omega$; in particular, for the case $\omega=1$, the
result for exponent $\gamma$ is significatively larger than
$\gamma=1/4$.

\begin{table}
\begin{tabular}{cccc}\hline\hline
\multicolumn{4}{c}{$\omega=0$} \\ \hline
    $\alpha=0$ & $\alpha=0.4$ & $\alpha=0.6$ & $\alpha=0.7$ \\
 $\gamma=0.247 \pm 0.003$ & $\gamma=0.248 \pm 0.003$ & $\gamma=0.248 \pm 0.001$ & $\gamma=0.245 \pm 0.003$ \\
    \hline \hline
 \multicolumn{4}{c}{$\alpha=1/2$} \\ \hline
 $\omega=0.2$ & $\omega=0.5$ & $\omega=1/\sqrt{2}$ & $\omega=1$ \\
 $\gamma=0.255 \pm 0.006$ & $\gamma=0.259 \pm 0.003$ & $\gamma=0.264 \pm 0.008$ & $\gamma=0.284 \pm 0.001$ \\ \hline\hline
\end{tabular}
\caption[]{Finite size scaling exponents $\gamma$ for the cases depicted in Fig. \ref{fig:exponentes}}
\label{tab:exponentes}
\end{table}

\subsection{Decoherence factor for the first order ESQPT}

For the case $\omega \neq 0$, the Hamiltonian considered produce, in
addition to the continuous ESQPT studied in the preceding subsection,
a first order ESQPT at energy  $E^{(1)}_c$ . This critical energy
can be estimated calculating the
local minima in the energy surface $\mathcal{H} (\phi,\xi)$, as it is
shown in Fig.  \ref{fig:energia}. Inserting this value in
Eq. (\ref{eq:lambdac}) a critical coupling $\lambda_c^{(1)}$ is
obtained. For the case $\alpha=1/2$ and $\omega=1/\sqrt{2}$ the first
order EQSPT is obtained at $\lambda_c^{(1)} \approx 1.05$.

\begin{figure}
\begin{center}
\rotatebox{-90}{\scalebox{0.25}[0.25]{\includegraphics{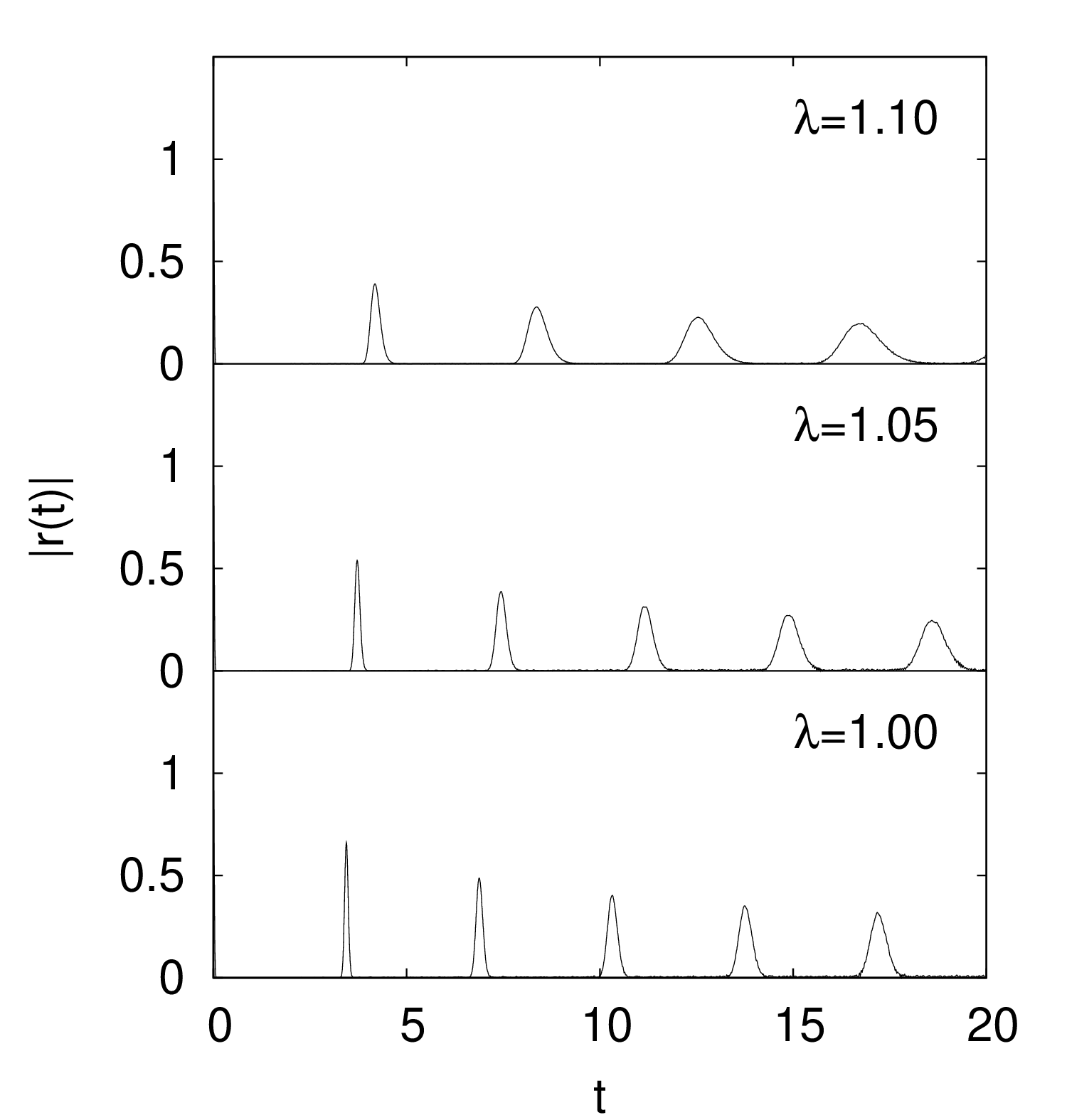}}} \caption[]{
$|r(t)|$ for $\alpha=1/2$, $\omega=1/\sqrt{2}$ and three different
  values of $\lambda$. In all cases $N=10000$.} \label{fig:1st}
\end{center}
\end{figure}

In Fig. \ref{fig:1st} we show the exact result for the decoherence
factor $| r(t) |$ for $\alpha=1/2$, $\omega=1/\sqrt{2}$, and three
different values of $\lambda$ around $\lambda=\lambda_c^{(1)} \approx
1.05$. The most significative result is that no trace of critical
phenomena are observed in $|r(t)|$ -- the shape of this magnitude is
smooth around $\lambda=\lambda_c^{(1)}$. Moreover, Fig. \ref{fig:rmax}
confirms that $r_{max} (\lambda)$ also behaves in a smooth an
size-independent way. The conclusion is, thus, that the first-order
ESQPT does not affect the decoherence induced in the central qubit.

\section{Summary and conclusions}

The decoherence induced in a single qubit by its interaction with the
environment, modelled as a scalar two-level boson model, is
studied. The environment presents a quantum phase transition from
symmetric to non-symetric phases at around $\alpha=4/5$, which can be
first order ($\omega \ne 0$) or second order ($\omega=0$). In the
non-symmetric phase, the environment also presents excited state
quantum phase transitions (ESQPTs): a second order one for any
$\omega$ value at $E_c^{(2)}=0$, and also a first order one for
$\omega \ne 0$ at an energy $E_c^{(1)}<0$. We have shown that the
second order ESQPT affects dramatically the decoherence factor which
goes rapidly to zero. A finite size scaling study shows that in that
case the decoherence factor goes to zero at the critical point
following a power law. On the other hand, the first order ESQPT does
not affect the decoherence of the central qubit.

We have also shown that a mean field treatment provides a good
description of the decoherence factor $r(t)$, except in the regions
around the critical points. Therefore, more sophisticated
approximations are needed to obtain an analytical description of the
critical behavior of $r(t)$, and, particulary, to estimate the
critical exponent $\lambda$.

\section*{Acknowledgements}

This work has been partially supported
by the Spanish Ministerio de Educaci\'on y Ciencia and by the European
regional development fund (FEDER) under projects number
FIS2008-04189,  FIS2006-12783-C03-01
FPA2006-13807-C02-02 and FPA2007-63074, by CPAN-Ingenio, by Comunidad de Madrid under
project 200650M012, CSIC and by Junta de
Analuc\'{\i}a under projects FQM160, FQM318, P05-FQM437 and
P07-FQM-02962. A.R. is supported by the Spanish program "Juan de la
Cierva" and P. P-F. is supported by a FPU grant of the Spanish
Ministerio de Educaci\'on y Ciencia.

\end{document}